\documentclass[aps,showpacs,floatfix]{revtex4}

\usepackage{amssymb}
\usepackage{amsmath}
\usepackage{epsfig}
\usepackage{subfigure}
\usepackage{multirow}

\begin{document}

\title{The Parameterised Post-Newtonian Limit of Fourth-Order Theories of Gravity}
\author{Timothy Clifton}
\email{TClifton@astro.ox.ac.uk}
\affiliation{Department of Astrophysics, University of Oxford, Oxford OX1 3RH, UK}
\date{\today }

\begin{abstract}

We determine the full post-Newtonian limit of theories of gravity
that extend general relativity by replacing the Ricci scalar, $R$, in the
generating Lagrangian by some analytic function, $f(R)$.  We restrict ourselves
to theories that admit Minkowski space as a suitable
background, and perform a perturbative expansion in the manner prescribed by the parameterised
post-Newtonian formalism.  Extra potentials are found to be present that are not
accounted for in the usual treatment, and a discussion is provided
on how they may be used to observationally distinguished these theories
from general relativity at the post-Newtonian level.

\end{abstract}

\maketitle

\section{Introduction}

There exists an extensive literature on relativistic theories that generalise
Einstein's theory of general relativity (GR), and that reduce to GR in
the appropriate limits.  A particularly appealing class
of these generalisations are the fourth-order theories.  These
are theories derived from a Lagrangian density that is a
scalar function of contractions of the Riemann tensor only.
Considerations of fourth-order theories of gravity have a long history,
having been first considered by Eddington in as early as the 1920's
\cite{Edd}.  One frequently considered generalisation is to replace the Ricci curvature scalar,
$R$, in the Einstein-Hilbert action with some analytic function, $f(R)$ (see e.g. \cite{buch, ker,
  bar, mag}).  It is the post-Newtonian limit of such theories that we
will be interested in here.

There are a variety of reasons why one may wish to consider these
generalised theories.  Strong motivation comes the from their
renormalization properties in the presence of matter
fields \cite{Stel}.  Other motivation can be found from cosmological
considerations where it has been found that generalising the
Einstein-Hilbert action can be of use for better understanding the
late-time acceleration of the universe \cite{tur, new}, early universe
inflation \cite{berk, brun, herv} and the nature of the initial
singularity \cite{brrw, clftn, dnsby, exact}.  Whatever the motivation for
considering generalised fourth-order theories, it is
essential that we sufficiently understand their weak-field limit, and
that they conform with the ever increasing body of observational data.

The usual frame-work for considering the weak-field effects of modified theories is
the parameterised post-Newtonian (PPN) formalism.  The first steps
towards this formalism were again made by Eddington \cite{Edd},
together with Robertson \cite{rob62} and Schiff \cite{sch67}, who
treated the planets as test bodies moving in the gravitational field
of the sun.  In their formalism they introduced the test metric
\begin{equation*}
ds^2=-\left(1-\frac{2m}{r}+\frac{2 \beta m^2}{r^2}\right) dt^2 +
\left( 1+\frac{2 \gamma m}{r} \right) d\bf{x}^2
\end{equation*}
where $d\bf{x}^2$ is the three dimensional Euclidean
line-element, $r$ is a radial coordinate and $m$ is the mass of the
central gravitating object.  The parameters $\beta$ and $\gamma$ are to be determined
by experiment.  By constraining them it is
possible to verify, or potentially disprove, GR and its
alternatives.  For example, GR predicts
$\beta=\gamma=1$, whilst Brans-Dicke theory \cite{Brans} predicts
$\beta=1$ and $\gamma=(1+\omega)/(2+\omega)$, where $\omega$ is the
Brans-Dicke coupling constant.  Using radio communications from the Cassini
spacecraft the constraint $\gamma = 1 + (2.1 \pm 2.3) \times 10^{-5}$
has been obtained \cite{Bert}.  This is evidently in good
agreement with GR, and can be used to constrain the
Brans-Dicke parameter to be $\omega\gtrsim 40 \; 000$, to $2\sigma$.  

Our goal here is to calculate the post-Newtonian limit of fourth-order
theories, so that they can be the subject of observation in a similar
manner.  To achieve this we will consider them in the context of the
full PPN formalism.  This formalism is a generalisation of the
Eddington, Robertson, Schiff parameterisation outlined above, and
allows for more general configurations of matter fields than a single
point-like gravitating source.  The PPN approach was developed
primarily by Nordvedt and Will \cite{Nor}, and is explained in detail
in \cite{tegp}.  We will give a brief explanation of the PPN approach
below, in so far as will be required for coherence of this work.  For
a more complete exposition the reader is referred to \cite{tegp}.

Absent in previous studies is a rigorous exemplification of how fourth-order
theories fit into the PPN formalism.  Here we will remedy this by extending the PPN
formalism to include fourth-order $f(R)$ theories.  In doing so we will
attempt to maintain, to the highest degree possible, the principles
and spirit of the PPN formalism, as expounded in \cite{tegp}.
Previous attempts have been made in this direction by Capozziello and
Troisi \cite{cap} and Olmo \cite{olmo}.  These authors attempt to
derive the post-Newtonian limit of $f(R)$ theories by appealing to
their equivalence with scalar-tensor theories \cite{frst}.  Here we
work directly with the fourth-order theory, and by direct integration
of the field equations find results that explicity state
their post-Newtonian limit.  In doing so we restrict our attention to
the subset of theories that admit Minkowski space as a suitable
background, and find that it is necessary to introduce a number of new
post-Newtonian potentials.  Theories with other backgrounds will
be investigated elsewhere.  We note that the post-Newtonian limit of
Gauss-Bonnet gravity has been found by Sotiriou and Barausse \cite{sot}.

In section \ref{4th} we introduce theories that are derivable from a
Lagrangian density of the form $\mathcal{L}=f(R)$.  The field
equations are derived, and a discussion is given of the condition that
Minkowski space be an appropriate background.  Section \ref{ppnsec} gives
a brief introduction to the PPN formalism, in as much as is
required for the self-consistency of this article.  In sections
\ref{newtonsec} and \ref{pnsec} we find the Newtonian and
post-Newtonian limits, respectively.  These calculations are performed in
the presence of a perfect fluid, and result in many new potentials
that are not usually present.  In section \ref{gaugesec} the results
found in the previous two sections are transformed into the standard
post-Newtonian gauge, in which the spatial part of the metric is
diagonal and derivatives of quantities associated with the matter
fields are removed.  Section \ref{discussion} gives a discussion of
the results obtained, and in particular gives the relevant
post-Newtonian limit if the usual Newtonian potential is to dominate
at the Newtonian level of approximation.  In section \ref{conclusions}
we conclude, and indicate the principle ways in which the $f(R)$
theories we are considering may be observationally distinguishable
from GR at the post-Newtonian level.  The appendices give some details
of the more lengthy calculations.

\section{Fourth-Order Theories}
\label{4th}

The Lagrangian density for the theories we will be considering is
\begin{equation}
\label{L}
\mathcal{L}= f(R).
\end{equation}
The action associated with this density is then given by
\begin{equation}
S = \int \sqrt{-g} \mathcal{L} + S_m,
\end{equation}
where $S_m$ denotes the action associated with the matter fields.  Extremizing this
action with respect to the metric results in the field equations
\begin{equation}
\label{field}
f^\prime R_{\mu \nu} -\frac{1}{2} f g_{\mu \nu}+{{f^\prime}_;}^{\sigma
  \rho} (g_{\mu \nu} g_{\sigma \rho}-g_{\mu \sigma} g_{\nu \rho}) = 8
  \pi T_{\mu \nu}
\end{equation}
where primes denote differentiation with respect to $R$ and $T_{\mu \nu}$
is the energy-momentum tensor, defined in terms of $S_m$ and $g_{\mu \nu}$ in the usual
way.  Throughout we us Greek letters to run over all space-time
indices and choose units so that $c=G=1$. The Lagrangian
formulation of these theories guarantees the covariant conservation of
energy-momentum.

In order to define a perturbative expansion we must first decide the
appropriate background to expand about.  In the usual PPN treatment
this background is taken to be Minkowski space, and the metric is then
expanded as
\begin{equation}
\label{h}
g_{\mu \nu}=\eta_{\mu \nu} + h_{\mu \nu}
\end{equation}
where $h_{\mu \nu} \ll 1$.  Such an expansion is well motivated in
GR where it is known that in the absence of any
matter fields Birkhoff's theorem ensures staticity of spherically
symmetric space-times, and that these space-times will be asymptotically
flat.  Adding small amounts of matter to such a background is then
well modelled by the perturbative expansion given by (\ref{h}), and
one need only be concerned with the effects of matching this
region to a suitable cosmological solution at large distances.

More care is required with fourth-order theories, where staticity and
asymptotic flatness cannot be so easily assumed.  Birkhoff's theorem
is not valid in these theories, and so staticity, if it
is required, must be imposed as an extra condition on spherically
symmetric vacua.  However, the imposition of this extra symmetry does
\textit{not}, in general, result in Minkowski space as a suitable background.  It was shown
explicitly in \cite{Power} that if any single power of the gravitational Lagrangian
dominates, other than the Einstein-Hilbert one, then
spherically symmetric vacuum space-times are not asymptotically
attracted to Minkowski form.  Instead, the line-element asymptotically approaches
\begin{equation*}
ds^2 \rightarrow -r^n dt^2 + dr^2 + r^2 d\Omega^2
\end{equation*}
as $r \rightarrow \infty$, where $n=0$ only if the Einstein-Hilbert term dominates.  Such
asymptotic behaviour can be readily shown to be incompatible with
observations, unless $n$ is very small \cite{Power}.

An alternative prescription to the imposition of a static background was
investigated in \cite{Cli06}.  Here the method of imposing time
independence, as outlined above, was contrasted with
the method of assuming asymptotic homogeneity and isotropy of the
background vacuum.  The assumption of homogeneity and isotropy removes
the need for $r$ dependence in the asymptotic form of the metric, but
at the expense of introducing time dependence.  It was shown in \cite{Cli06}, using
exact solutions as well as perturbative expansions, that the choice of
symmetries for the background has a demonstrable effect on the
weak-field expansions that are performed within them.

In short, one must make a choice of background to expand about, and
this choice can have important consequences for the expansion itself.
If any single power of the gravitational Lagrangian other than the
Einstein-Hilbert one dominates at asymptotically large distances, then
Minkowski space may not be an appropriate, stable choice of
background.  One is then forced to recant either the time independence
or the homogeneity of the background metric, with non-trivial consequences.

Here we will avoid these difficulties by considering only theories in
which the Einstein-Hilbert term dominates in the low curvature regime.
We are then justified in performing a perturbative expansion about
Minkowski space, which allows a more direct comparison with the
usual PPN approach.  The post-Newtonian limit of other $f(R)$ theories will
be dealt with in a future study.  Fourth-order theories that
admit a Minkowski background, and are an analytic functions of $R$, can
then be written as
\begin{equation}
\label{f}
f(R) = \sum_{i=1}^{\infty} c_i R^i
\end{equation}
where the $c_i$ are a set of real, positive valued constants.
The post-Newtonian analysis that
is to follow will show how the $c_i$ are manifest in the
weak-field limit, and hence how they can be potentially observed with
gravitational experiments and observations.

\section{The PPN Approach}
\label{ppnsec}

This section is a recapitulation of the PPN formalism, as propounded in
\cite{tegp}, and as is necessary for coherence of this article.
The PPN formalism is a perturbative treatment of weak-field gravity.
Such an expansion requires a small parameter to expand in.  An ``order
of smallness'' is therefore defined by
\begin{equation*}
U \sim v^2\sim\frac{p}{\rho} \sim \Pi \sim O(2)
\end{equation*}
where $U$ is the Newtonian potential, $v$ is the velocity a fluid
element, $p$ is the pressure of the fluid, $\rho$ is its rest-mass
density and $\Pi$ is the ratio of energy density to rest-mass
density.  Time derivatives are also taken to have an order of
smallness associated with them, relative to spatial derivatives:
\begin{equation*}
\frac{\vert \partial/\partial t \vert}{\vert \partial/\partial x
  \vert} \sim O(1).
\end{equation*}
(Recall that we have chosen to set $c=1$).  The PPN formalism now
proceeds as an expansion in this order of smallness.

The equations of motion show that for time-like particles
propagating along geodesics the level of approximation
required to recover the Newtonian limit is $g_{00}$ to $O(2)$, with no
other knowledge of the metric components beyond the background level being necessary.  The
post-Newtonian limit for time-like particles requires a knowledge of
\begin{align*}
&g_{00} \qquad \text{to} \qquad O(4)\\
&g_{0i} \qquad \text{to} \qquad O(3)\\
&g_{ij} \qquad \text{to} \qquad O(2).
\end{align*}
Latin letters are used to denote spatial indices.  To obtain the
Newtonian limit of null particles we only need to
know the metric to background order:  Light follows straight lines, to
Newtonian accuracy.
The post-Newtonian limit of null particles requires a knowledge of
$g_{00}$ and $g_{ij}$ both to $O(2)$.

Using the expansion (\ref{h}) we can now calculate the Ricci and
energy-momentum tensors to the appropriate orders.  However, before
doing so it is worth recognising that we have four gauge freedoms,
associated with four coordinate choices.  Specifying the four gauge
conditions
\begin{align}
\label{gauge1}
&h_{i0,i}=\frac{1}{2} h_{ii,0} + O(5)\\
\label{gauge2}
&h_{ij,j}=\frac{1}{2} h_{jj,i} -\frac{1}{2} h_{00,j} +O(4)
\end{align}
allows the components of the Ricci tensor to be written
\begin{align}
\label{R00}
&R_{00}= -\frac{1}{2} \nabla^2 h_{00} -\frac{1}{2} \vert \nabla h_{00}
  \vert^2+ \frac{1}{2} h_{jk} h_{00,jk} +O(6)\\
\label{R0i}
&R_{0i}= -\frac{1}{2} \nabla^2 h_{0i} -\frac{1}{4}h_{00,i0} +O(5)\\
\label{Rij}
&R_{ij}= -\frac{1}{2} \nabla^2 h_{ij} +O(4)
\end{align}
where $\nabla^2=\partial^i \partial_i$ is the Laplacian on three
dimensional Euclidean space.  We still have the freedom to make gauge
transformations of the form
\begin{equation*}
x^{\mu} \rightarrow x^{\mu} + \xi^{\mu}
\end{equation*}
and we will use this freedom in the following analysis to transform
to a ``standard post-Newtonian gauge'' in which the spatial part of
the metric is diagonal, and terms containing time derivatives are
removed.  The components of the stress-energy tensor, to the relevant order, are
\begin{align}
\label{T00}
&T_{00} = \rho (1+\Pi +v^2 - h_{00})\\
\label{T0i}
&T_{0i} = -\rho v_i\\
\label{Tij}
&T_{ij} = \rho v_i v_j +p \delta_{ij}.
\end{align}
We can now substitute these expressions for $R_{\mu\nu}$ and $T_{\mu\nu}$ into the field
equations (\ref{field}), together with (\ref{f}), and solve the
equations order by order in perturbations.  Transforming to an
appropriate gauge will then yield the PPN limit of these fourth-order theories.

\section{The Newtonian Limit}
\label{newtonsec}

The Newtonian limit of these theories will now be investigated.  This
limit has been found before for a point-like mass at the origin,
originally by \cite{Pech}, and again several times since.  Here we
find the general solution for a space-time containing a perfect fluid.

Beginning with the trace of the field equations (\ref{field}), we have
to $O(2)$
\begin{equation}
\label{trace}
\nabla^2 R^{(2)} - \frac{c_1}{6 c_2} R^{(2)} = -\frac{4 \pi}{3 c_2} \rho
\end{equation}
where $R^{(2)}$ denotes the Ricci scalar to $O(2)$.  This is an
inhomogeneous Helmholtz equation which has the solution
\begin{equation}
\label{R}
R^{(2)} = \frac{1}{3 c_2} \int \frac{\rho(\bf{x}^\prime)}{\vert
  \bf{x}-\bf{x}^\prime \vert} e^{-\sqrt{\frac{c_1}{6 c_2}} \vert
  \bf{x}-\bf{x}^\prime \vert} d^3x^\prime.
\end{equation}
We have ignored here the other possible root in the exponential, which is an
equally valid solution of the Helmholtz equation, but does not give an
appropriate limit at asymptotically large distances.  It can be seen
that (\ref{R}) is an exponentially decaying Yukawa potential if $c_1$ and
$c_2$ have the same signs, and is a damped oscillatory function if
they have opposite signs.

The $0-0$ and trace field equations, given by (\ref{field}), can now be written to $O(2)$ as
\begin{equation*}
\nabla^2 \left( \frac{1}{4} c_1 h^{(2)}_{00} +\frac{1}{4} c_1
h^{(2)}_{ii}+2 c_2 R^{(2)} \right) = -8\pi \rho.
\end{equation*}
and
\begin{equation*}
\nabla^2 \left(h^{(2)}_{ii} +5 h^{(2)}_{00} \right) = - \frac{64
  \pi}{c_1} \rho
\end{equation*}
where the $O(2)$ parts of (\ref{R00}) and (\ref{T00}) have been
used.  Solving these two inhomogeneous Poisson equations
simultaneously we find
\begin{equation}
\label{h002}
h^{(2)}_{00}=\frac{2}{c_1} (U+c_2 R^{(2)})
\end{equation}
where
\begin{equation}
\label{U}
U \equiv \int \frac{\rho(\bf{x}^\prime)}{\vert \bf{x}-\bf{x}^\prime \vert}
d^3x^\prime
\end{equation}
is the Newtonian potential.  From (\ref{h002}) it can be seen that the Ricci scalar
itself acts as a Newtonian level potential.

\section{The Post-Newtonian Limit}
\label{pnsec}

In this section we investigate the Post-Newtonian limit of the these
theories.  Expressions for $h_{ij}$ to $O(2)$, $h_{0i}$
to $O(3)$ and $h_{00}$ to $O(4)$ are found.  These expressions will not
be in the standard post-Newtonian gauge, discussed above.  We
perform a transformation into this gauge in the subsequent section.

\subsection{The $h_{ij}$ terms}

We begin by evaluating the terms $h_{ij}$ to $O(2)$.  These
quantities, together with (\ref{h002}) above, are sufficient to
determine the post-Newtonian limit of null geodesics.

The $i-j$ field equation (\ref{field}) can now be written
\begin{equation*}
\nabla^2 \left(\frac{c_1}{2} h_{ij}^{(2)} + c_2 \delta_{ij} R^{(2)} + 12
\frac{c_2^2}{c_1} R_{,ij}^{(2)} -4 \frac{c_2}{c_1} U_{,ij} \right) = -4 \pi
\rho \delta_{ij}
\end{equation*}
where we have made use of the expressions above for $R_{ij}$ and
$T_{ij}$, (\ref{Rij}) and (\ref{Tij}), the trace equation
(\ref{trace}) and the definition of $U$, (\ref{U}).  This
equation can be integrated to give
\begin{equation}
\label{hij2}
h_{ij}^{(2)} = \frac{2}{c_1} \left( U \delta_{ij}- c_2 \delta_{ij}
R^{(2)}-12 \frac{c_2^2}{c_1} R_{,ij}^{(2)} + 4 \frac{c_2}{c_1} U_{,ij}
\right).
\end{equation}
Equation (\ref{hij2}) is not diagonal, and so is not in the standard
post-Newtonian gauge.  In the next section will remove the
off-diagonal components with the appropriate transformation.

\subsection{The $h_{0i}$ terms}

As discussed above, the first non-zero contribution to the $h_{0i}$
terms is at $O(3)$.  The $0-i$ field equation (\ref{field}) can now be
written
\begin{equation*}
\nabla^2 \left( c_1 h_{0i}^{(3)} +30 \frac{c_2^2}{c_1} R^{(2)}_{,0i} -
10 \frac{c_2}{c_1} U_{,0i} -\frac{1}{2} V_i + \frac{1}{2}
W_i  \right) = 16 \pi \rho v_i
\end{equation*}
where we have used the expressions for $R_{0i}$ and $T_{0i}$,
(\ref{R0i}) and (\ref{T0i}), the expression for $h_{00}$ to $O(2)$,
(\ref{h002}), the trace equation, (\ref{trace}), and the two new
potentials $V_i$ and $W_i$, that are defined as
\begin{align}
\label{V}
V_i &\equiv \int \frac{\rho(\mathbf{x}^\prime) v_i(\bf{x}^\prime)}{\vert
  \bf{x}-\bf{x}^\prime \vert} d^3x^\prime\\
\label{W}
W_i &\equiv \int \frac{\rho(\mathbf{x}^\prime) (\mathbf{v}(\mathbf{x}^\prime) \cdot
  (\mathbf{x}-\mathbf{x}^\prime)) (x-x^\prime)_i}{\vert \mathbf{x}-\mathbf{x}^\prime
  \vert^3} d^3x^\prime
\end{align}
as in the usual PPN treatment, so that $\nabla^2 (W_i-V_i) = 2
U_{,0i}$.  Use has also been made of the conservation equation
\begin{equation}
\label{fluid}
\frac{\partial \rho}{\partial t} + \nabla \cdot (\rho \mathbf{v}) = 0.
\end{equation}
Integrating the field equation above we now have
\begin{equation}
\label{h0i3}
h_{0i}^{(3)} = -\frac{7}{2 c_1} V_i - \frac{1}{2 c_1} W_i + 10
\frac{c_2}{c_1^2} U_{,0i} - 30 \frac{c_2^2}{c_1^2} R_{,0i}^{(2)}
\end{equation}
which, again, will be subject to a gauge transformation in the
next section.

\subsection{The $h_{00}$ term to $O(4)$}

At this order of perturbation the equations become more unsightly, and so we choose to relegate the majority
of them to appendices, stating here only the results.  Hence,
in the gauge specified by (\ref{gauge1}) and
(\ref{gauge2}), the $h_{00}$ term to $O(4)$ is
\begin{align}
\nonumber
h^{(4)}_{00} = &-\frac{2}{c_1^2} U^2 + 2 \frac{c_2^2}{c_1^2} R^2 - \frac{16 c_2}{3 c_1^2} U R - 36
\frac{c_2^2}{c_1^2} R_{,00} +12 \frac{c_2}{c_1^2} U_{,00}
+ 8 \frac{c_2}{c_1^3} \vert \nabla U \vert^2 - 24 \frac{c_2^3}{c_1^3}
\vert \nabla R \vert^2 - 16 \frac{c_2^2}{c_1^3} \nabla U \cdot \nabla R
\\ \nonumber &- \frac{7}{18 \pi c_1} \mathcal{V}(UR) + \frac{3 c_2}{4 \pi
c_1} \mathcal{V}(R^2) + \frac{64}{9 c_1^2} \mathcal{V}(\rho U) - \frac{44 c_2}{3 c_1^2} \mathcal{V}( \rho R)-
\frac{40 c_2}{3 c_1^3} \mathcal{V}(\nabla \rho \cdot \nabla U) + \frac{40
  c_2^2}{c_1^3} \mathcal{V}(\nabla \rho \cdot \nabla R) \\ \nonumber &+ \frac{2}{c_1} \mathcal{V}(\rho
\Pi) + \frac{4}{c_1} \mathcal{V}(\rho v^2) + \frac{6}{c_1} \mathcal{V}(p) - \frac{1}{4\pi} \left( \frac{c_2}{c_1}
- \frac{c_3}{2 c_2} \right)X(R^2) + \frac{1}{6 \pi c_1} X(UR)- \frac{4}{3 c_1^2} X(\rho U)
+ \frac{8 c_2}{3 c_1^2} X( \rho R)\\ &+ \frac{8 c_2}{3 c_1^3} X(\nabla \rho \cdot \nabla U)  
 \label{h004} - \frac{8 c_2^2}{c_1^3} X(\nabla \rho \cdot \nabla R)-\frac{2}{c_1} X(p)+ \frac{2}{3
   c_1}  X(\rho \Pi)-\sqrt{\frac{2 c_2}{3 c_1^3}} \hat{\chi}_{,00}.
\end{align}
The derivation of this result can be found in appendix
\ref{appendixA}, where it is given there as equations (\ref{R4}) and
(\ref{h004a}).  The new potentials $\mathcal{V}$, $X$ and $\hat{\chi}$ are
defined as
\begin{align}
\label{VX2}
\mathcal{V}(Q) &\equiv \int \frac{Q^\prime}{\vert \mathbf{x}-\mathbf{x^\prime}
  \vert} dx^{\prime 3}\\
\label{WX2}
X(Q) &\equiv \int \frac{Q^\prime e^{-\sqrt{\frac{c_1}{6 c_2}} \vert
  \bf{x}-\bf{x}^\prime \vert}}{\vert \mathbf{x}-\mathbf{x}^\prime
  \vert} d^3x^\prime\\
\label{chi2}
\hat{\chi} &\equiv \int \rho^\prime e^{-\sqrt{\frac{c_1}{6 c_2}} \vert
  \bf{x}-\bf{x}^\prime \vert} d^3x^\prime.
\end{align}
Primes here label quantities that are functions of
$\mathbf{x^\prime}$.  It should be noted that this definition of
$\mathcal{V}$ is degenerate with some of the usual PPN parameters:  For
example, $\mathcal{V}(\rho v^2)$ is identical to the potential $\Phi_2$ of
\cite{tegp}.  We use this definition of $\mathcal{V}$ as it is
convenient for expressing the new potentials.

Again, this result is not in the standard post-Newtonian gauge.  In
the following section we will transform it so that the terms
proportional to $R_{,00}$, $U_{,00}$, $\vert \nabla U \vert^2$,
$\vert \nabla R \vert^2$, $\nabla U \cdot \nabla R$  and
$\hat{\chi}_{,00}$ are eliminated.

\section{Gauge Transforming}
\label{gaugesec}

In the preceding section we used the gauge specified by
conditions (\ref{gauge1}) and (\ref{gauge2}).  This has been a
convenient choice, and has allowed integration of the field equations
to post-Newtonian accuracy.  However, it is desirable to transform the
results found above to a gauge in which the spatial part of the metric
is diagonal, and in which the metric takes it simplest form.  By making
the coordinate transformation $x^{\mu} \rightarrow x^{\mu} + \xi^{\mu}$
the metric is transformed in such a way that
\begin{equation*}
h_{\mu \nu} \rightarrow h_{\mu \nu} - \xi_{\mu ; \nu} - \xi_{\nu ;
  \mu} + O(\xi^2).
\end{equation*}
Then by making the choices
\begin{align*}
\xi_0 &= 6 \frac{c_2}{c_1^2} U_{,0} - 18 \frac{c_2^2}{c_1^2} R_{,0} -
\sqrt{\frac{c_2}{6 c_1^3}} \hat{\chi}_{,0}\\
\xi_i &= 4 \frac{c_2}{c_1^2} U_{,i} - 12 \frac{c_2^2}{c_1^2} R_{,i}
\end{align*}
the metric perturbations transform as
\begin{align*}
h_{ij}^{(2)} &\rightarrow h_{ij}^{(2)} +24 \frac{c_2^2}{c_1^2} R_{,ij}
-8 \frac{c_2}{c_1^2} U_{,ij}\\
h_{0i}^{(3)} &\rightarrow h_{0i}^{(3)} -10 \frac{c_2}{c_1^2} U_{,0i}
+30 \frac{c_2^2}{c_1^2} R_{,0i} + \sqrt{\frac{c_2}{6 c_1^3}}
\hat{\chi}_{,0i}\\
h_{00}^{(4)} &\rightarrow h_{00}^{(4)} - 12 \frac{c_2}{c_1^2} U_{,00}
+36 \frac{c_2^2}{c_1^2} R_{,00}+ \sqrt{\frac{2 c_2}{3 c_1^3}}
\hat{\chi}_{,00} - 8 \frac{c_2}{c_1^3} \vert \nabla U \vert^2 + 24
\frac{c_2^3}{c_1^3} \vert \nabla R \vert^2 +16 \frac{c_2^2}{c_1^3}
\nabla U \cdot \nabla R
\end{align*}
whilst $h_{00}^{(2)}$ is unchanged.  These transformations are exactly
what is required to diagonalize the spatial part of the
metric, and to remove unwanted terms from the other metric
components.  The final form of the perturbed metric can now be
written to the required order as
\begin{align}
\nonumber
g_{00} = &-1 + \frac{2}{c_1} \left(U+ c_2 R \right)
-\frac{2}{c_1^2} U^2 + 2 \frac{c_2^2}{c_1^2} R^2 - \frac{16 c_2}{3
  c_1^2} U R - \frac{7}{18 \pi c_1} \mathcal{V}(UR) + \frac{3 c_2}{4 \pi c_1} \mathcal{V}(R^2) +
\frac{64}{9 c_1^2} \mathcal{V}(\rho U) \\ \nonumber &- \frac{44 c_2}{3 c_1^2} \mathcal{V}( \rho R)-
\frac{40 c_2}{3 c_1^3} \mathcal{V}(\nabla \rho \cdot \nabla U) + \frac{40
  c_2^2}{c_1^3} \mathcal{V}(\nabla \rho \cdot \nabla R) + \frac{2}{c_1} \mathcal{V}(\rho
\Pi) + \frac{4}{c_1} \mathcal{V}(\rho v^2) + \frac{6}{c_1} \mathcal{V}(p) \\ \nonumber &+ \frac{1}{6 \pi
  c_1} X(UR) - \frac{1}{4\pi} \left( \frac{c_2}{c_1}
- \frac{c_3}{2 c_2} \right)X(R^2)
- \frac{4}{3 c_1^2} X(\rho U)+
\frac{8 c_2}{3 c_1^2}X( \rho R) + \frac{8 c_2}{3 c_1^3}X(\nabla \rho \cdot \nabla U)  
 \\ \label{g00} &- \frac{8 c_2^2}{c_1^3}X(\nabla \rho
\cdot \nabla R) -\frac{2}{c_1} X(p)+ \frac{2}{3c_1}X(\rho \Pi)\\ \label{g0i}
g_{0i} = &-\frac{7 V_i}{2 c_1} - \frac{W_i}{2 c_1} + \frac{X(\rho v_i)}{6 c_1}
 - \frac{Y_i}{6 c_1} -\frac{Z_i}{6\sqrt{6 c_1 c_2}}
\\ \label{gij}
g_{ij} = &\left(1+\frac{2}{c_1} \left(U -c_2 R \right) \right) \delta_{ij}
\end{align}
where we have introduced the new potentials $Y_i$ and $Z_i$,
which are defined as
\begin{align}
Y_i &\equiv \int \frac{\rho^\prime \mathbf{v}^\prime \cdot
  (\mathbf{x}-\mathbf{x}^\prime) (x-x^\prime)_i}{\vert
  \mathbf{x}-\mathbf{x}^\prime \vert^3 } e^{-\sqrt{\frac{c_1}{6 c_2}} \vert
  \bf{x}-\bf{x}^\prime \vert} d^3x^\prime\\
Z_i &\equiv \int \frac{\rho^\prime \mathbf{v}^\prime \cdot
  (\mathbf{x}-\mathbf{x}^\prime) (x-x^\prime)_i}{\vert
  \mathbf{x}-\mathbf{x}^\prime \vert^2 } e^{-\sqrt{\frac{c_1}{6 c_2}} \vert
  \bf{x}-\bf{x}^\prime \vert} d^3x^\prime,
\end{align}
and where use has again been made of the conservation equation
(\ref{fluid}).  The reader will notice in equation (\ref{g00})
potentials that are functions of gradients of $\rho$, $U$ and $R$,
such as $\mathcal{V}(\nabla \rho \cdot \nabla U)$.  This type of term
is not of the usual PPN form, where the metric contains functionals of
rest mass, energy, pressure and velocity, but not their
gradients \cite{tegp}.  In appendix \ref{appendixB} we re-express the offending terms in a
more proper PPN form, where gradients are absent.

\section{Discussion}
\label{discussion}

We have found the PPN limit of analytic $f(R)$ theories of
gravity that allow an asymptotically Minkowski background.  The
weak-field metric for these theories, in the presence of a perfect fluid and in
the standard post-Newtonian gauge, is given by equations (\ref{g00}),
(\ref{g0i}) and (\ref{gij}).  We shall now proceed to investigate their
form.

Firstly, we will consider the limit where the higher order
contributions to the action are vanishing, so that $c_3 \rightarrow c_2
\rightarrow 0$.  In such a limit it can be seen from the trace
equation, (\ref{trace}), that $R \rightarrow 8 \pi \rho/c_1,$
and that the metric specified by (\ref{g00}), (\ref{g0i}) and (\ref{gij})
then reduces to
\begin{align*}
g_{00} \rightarrow &-1 + 2 U -2 U^2+
4 \mathcal{V}(\rho U) + 2 \mathcal{V}(\rho
\Pi) + 4 \mathcal{V}(\rho v^2) + 6
\mathcal{V}(p)
\\
g_{0i} \rightarrow &-\frac{7}{2} V_i - \frac{1}{2} W_i 
\\
g_{ij} \rightarrow &\left(1+2 U \right) \delta_{ij}
\end{align*}
where the two terms $\mathcal{V}(\rho U)$ and $\mathcal{V}(R U)$ in
(\ref{g00}) have contributed to $\mathcal{V}(\rho U)$ in the
expression above.  Here we have set $c_1=1$.  This metric is
the PPN limit of GR (see \cite{tegp} for details).

In order to be considered viable for non-zero $c_2$ and $c_3$ it is necessary for these theories to
reduce to Newtonian gravity in the appropriate limit.  From the $O(2)$
term of the $g_{00}$ component of the metric, (\ref{g00}), it can be
seen that this can occur iff either
$$(i) \; 3 c_2 R \sim U \qquad \qquad \text{or} \qquad \qquad (ii)
\; 3 c_2 R \sim 0$$ on
observable length scales.  Condition $(i)$ is met if $$\vert \sqrt{c_1/6c_2} \vert L \ll 1$$ for the largest
length-scales on which Newtonian gravity has been observed, $L$.
Alternatively, condition $(ii)$ can be met if $R$ is a
decaying exponential with $$\sqrt{c_1/6c_2} l \gg 1,$$ where $l$ is
the smallest length scale on which Newtonian gravity has been
observed.  In both cases we can now find simple expressions for the post-Newtonian
limit, which will allow us to relate our results to the relevant observations. 

Let us first consider case $(i)$.  This case can be easily dismissed
by considering only the Newtonian limit of $g_{00}$ and
the post-Newtonian limit of $g_{ij}$, equations (\ref{g00}) and
(\ref{gij}) above.  If $\sqrt{c_1/6c_2} \vert \mathbf{x}
-\mathbf{x}^\prime \vert \ll 1$ then $3 c_2 R \simeq U$, and we must set $c_1=4/3$ to obtain
the appropriate Newtonian term $$g_{00} = -1+2 U+O(4).$$  However,
substitution of this value of $c_1$ into (\ref{gij}) in the same limit
gives $$g_{ij}= (1+U) \delta_{ij} +O(4),$$  as previously found by
Chiba, Smith and Erickcek \cite{Chiba}.  This result is entirely incompatible
with many observations of null geodesics (see e.g. \cite{Bert}) and so
we will not consider this case any further.

Let us now consider case $(ii)$.  Here the potentials containing
exponentials are expected to be sub-dominant to those without, on
observable length scales, so that the $R$ potential in $g_{00}^{(2)}$
is only effective at very small distances.  Setting $c_1=1$, and
discarding potentials which are exponentially suppressed with regards
to others, we then find that the metric in this case is given to the appropriate 
order as
\begin{align}
g_{00} \simeq &-1 + 2U - 2 U^2  + \frac{64}{9} \mathcal{V}(\rho U)  -
\frac{7}{18 \pi} \mathcal{V}(UR) + 2 \mathcal{V}(\rho 
\Pi) + 4 \mathcal{V}(\rho v^2) \nonumber \\ &\qquad \qquad \qquad
\qquad \qquad \qquad \qquad \qquad + 6 \mathcal{V}(p) -\frac{40 c_2}{3}\psi_1 + \frac{40}{3}
\sqrt{\frac{c_2}{6}} \psi_2+\frac{c_3}{8 \pi
  c_2} X(R^2) \label{g00f}
\\ 
\label{g0if}
g_{0i} \simeq &-\frac{7}{2} V_i- \frac{1}{2}W_i-\frac{Z_i}{6\sqrt{6 c_2}}
\\ 
\label{gijf}
g_{ij} \simeq &\left( 1+2 U\right) \delta_{ij}
\end{align}
where $\psi_1$ and $\psi_2$ are defined, as in appendix
\ref{appendixB}, by
\begin{align}
\psi_1 &\equiv \int \frac{\rho^\prime \rho^{\prime \prime} (\mathbf{x}-\mathbf{x}^\prime) \cdot
  (\mathbf{x}^\prime-\mathbf{x}^{\prime\prime})}{\vert
  \mathbf{x}-\mathbf{x}^\prime \vert^3 \vert
  \mathbf{x}^\prime-\mathbf{x}^{\prime\prime} \vert^3} d^3x^\prime
  d^3x^{\prime \prime}\\
\psi_2 &\equiv \int \frac{\rho^\prime \rho^{\prime \prime} (\mathbf{x}-\mathbf{x}^\prime) \cdot
  (\mathbf{x}^\prime-\mathbf{x}^{\prime\prime})}{\vert
  \mathbf{x}-\mathbf{x}^\prime \vert^3 \vert
  \mathbf{x}^\prime-\mathbf{x}^{\prime\prime} \vert^2} e^{-\sqrt{\frac{c_1}{6 c_2}} \vert
  \bf{x}^\prime-\bf{x}^{\prime\prime} \vert} d^3x^\prime
  d^3x^{\prime \prime}.
\end{align}
In the metric above we have retained the term proportional to $Z_i$ in
the $g_{0i}$ components.  Although this term contains an exponential
suppression factor, it is of a different form to $V_i$ and
$W_i$ and so cannot necessarily be assumed to be negligibly small in
comparison to them.

Another significant difference in the metric above is that the
coefficient of the term $\mathcal{V}(\rho U)$ is $64/9$, instead of
its usual value of $4$ in GR, and the inclusion of the new potential
$\mathcal{V}(UR)$.  We have already seen that
in the limit $c_2 \rightarrow 0$ that the $\mathcal{V}(UR)$ term reduces to
$8 \pi \mathcal{V}(\rho U)/c_1$, which is exactly sufficient to
recover the GR limit of $4 \mathcal{V}(\rho U)$ in $g_{00}$.  The fact
that this potential is significant in the limit of small $c_2$, even
though it contains an exponential suppression term, is due
to the factor of $c_2$ in its denominator, as $R$ is present
in the integrand.  This causes it to approach a finite value, instead of
zero, when $c_2$ is small.  This potential cannot therefore be
considered negligible, as it is not necessarily exponentially smaller
than any other.

There are three further potentials in $g_{00}$ that are not
present in GR:  $\psi_1$, $\psi_2$ and $X(R^2)$.  The $\psi_1$ term contains no exponential suppression
factor, while the $\psi_2$ and $X(R^2)$ terms do.  However, as before, these
potentials are included none the less as they are not directly
suppressed with respect to any other.  The existence of $\psi_1$ in
the post-Newtonian limit is of particular interest as it is the first
new potential that is not exponentially suppressed, with obvious
significance for constraining the theory with observations.  The
$X(R^2)$ term is also of special interest as it is the only term with
a dependence on $c_3$.  In the limit $c_2 \rightarrow 0$
this term reduces to $$\frac{c_3}{8\pi c_2} X(R^2) \rightarrow 3 c_3
\rho^2,$$ which is non-zero when $c_3$ and $\rho \neq 0$.  This term
therefore provides an opportunity to test for deviations from GR at
the level of $R^3$ in the generating Lagrangian.

One may now wish to compare the metric obtained to the standard PPN metric, and
to read off the relevant parameters.  The standard PPN metric is given in the
present notation by
\begin{align}
g_{00}^{(PPN)} &= -1 +2 U -2 \beta U^2 +(2 \gamma +2 +\zeta_1)
\mathcal{V}(\rho v^2) \nonumber \\ &\qquad \qquad \qquad \qquad + 2 (3 \gamma-2 \beta+1+\zeta_2) \mathcal{V}
(\rho U) +2 (1+\zeta_3) \mathcal{V}(\rho \Pi) +6 (\gamma + \zeta_4)
\mathcal{V}(p)\label{g00ppn}\\
g_{0i}^{(PPN)} &= -\frac{1}{2} (4 \gamma +3 +\zeta_1) V_i - \frac{1}{2}
(1-\zeta_1) W_i\label{g0ippn}\\ \label{gijppn}
g_{ij}^{(PPN)} &= (1+2 \gamma U) \delta_{ij}
\end{align}
where $\beta$, $\gamma$ and $\zeta_i$ are the post-Newtonian
parameters, to be set for a particular gravitational theory.  We
have excluded here the preferred location and preferred frame terms, as they
are of no relevance for the present study.  Comparison of the metric
(\ref{g00f}), (\ref{g0if}) and (\ref{gijf}) with the above allows one
to read off the following values
\begin{equation*}
\beta = 1, \;\; \gamma =1, \;\; \zeta_1 =0, \;\; \zeta_3 =0 \qquad
\text{and} \qquad \zeta_4 =0,\\
\end{equation*}
as in GR.  The value of $\zeta_2$ is not so straightforwardly
determined, and a naive comparison would yield the result $\zeta_2 =
14/9$ instead of the usual value of zero in GR.  However, we have seen
above that the $\mathcal{V}( UR)$ term approaches $\mathcal{V}(\rho
U)$ for small $c_2$, and gives the GR result in the limit.  Care must
therefore be taken with the value of this parameter, and in the
present case it seems more appropriate to consider two contributions
towards $\zeta_2$ - one coming from the usual $\mathcal{V}(\rho U)$
term, and the other coming from $\mathcal{V}( UR)$.  The terms in
(\ref{g00f}) and (\ref{g0if}) proportional to $\psi_1$, $\psi_2$,
$X(R^2)$ and $Z_i$ are also inadequately accounted for in the PPN metric above.
Clearly, new terms are required if these potentials are to be included.

\section{Conclusions}
\label{conclusions}

We have determined here the post-Newtonian limit of
fourth-order theories of gravity that are analytic functions of the
Ricci tensor, and that admit Minkowski space as a
background.  In the Newtonian limit we have recovered the well known result that an
exponentially suppressed Yukawa potential is present.  These deviations
from Newton's law should be expected to be observed at small distance
scales, and a number of experimental efforts have been made to find them
(see e.g. \cite{Hoyle}).  These searches have not yet detected any
deviations from Newton's law at small distances, and so we must
consider terms containing exponential factors to be heavily
suppressed.

To determine the post-Newtonian limit of null geodesics we require
knowledge of the $g_{ij}$ components of the metric to an accuracy of
$O(2)$, as is given in equation (\ref{gij}).  We again find the
well known result that the only correction to this term at the
post-Newtonian level of accuracy is in the form of a Yukawa potential,
which must be heavily suppressed.  Comparison with the PPN metric
(\ref{gijppn}) then gives us that we should expect $\gamma =1$ for
these theories, and hence that experiments involving observations of
null geodesics should be unable to distinguish them from GR.

We then proceeded to determine the full post-Newtonian limit to $O(4)$
in the $g_{00}$ component, (\ref{g00}), and to $O(3)$ in the $g_{0i}$
component, (\ref{g0i}).  It is found that there exists a large number of
new potentials at this order of perturbations.  Using our knowledge that
terms containing exponential factors are heavily suppressed it is
possible to neglect a number of these potentials, resulting in
equations (\ref{g00f}) and (\ref{g0if}).  A comparison of these
results with (\ref{g00ppn}) and (\ref{g0ippn}) shows that the PPN
parameters $\beta$, $\zeta_1$, $\zeta_3$ and $\zeta_4$ all take the
same values in these $f(R)$ theories as they do in GR.  Experiments
which are designed to determine these parameters will therefore be
unable to distinguish between the two.  However, there are differences
between (\ref{g00f}) and (\ref{g0if}), and the PPN limit of GR, that may
be potentially observable.

A potentially significant difference with GR is the value of the PPN
parameter $\zeta_2$.  A direct comparison of (\ref{g00f}) with
(\ref{g00ppn}) appears to yield the result $\zeta_2=14/9$, which is a
significant difference from its value of zero in GR.  However, we know
that in the limit that GR is approached the new potential $\mathcal{V}(UR)$ in $g_{00}$
makes a contribution that is exactly enough to cancel the value of
$14/9$ above.  This strongly suggests that this potential should be
included when observational constraints are applied.   
The parameter $\zeta_2$ is usually associated with
violations of momentum conservation, and the (lack of)
self-acceleration of the binary pulsar PSR 1913+16 has led to the
bound $\zeta_2 < 4 \times 10^{-5}$ \cite{Willpsr}.  Furthermore, it
appears likely that observations of the binary system PSR J1738+0333
will offer even tighter constraints \cite{psr}.  However, these
previously obtained constraints may not
be directly applicable to the current theory.
Firstly, we know that $f(R)$ theories of gravity covariantly conserve
four-momentum exactly (due to their Lagrangian formulation).
Secondly, these constraints have been imposed in the absence of the
$\mathcal{V}(UR)$ potential.  The extent to which these systems
are able to offer constraints on $f(R)$ theories when the $\mathcal{V}(UR)$ term
is included remains to be determined.

Further opportunity for observationally constraining these theories
comes from the other extra potentials in (\ref{g00f}) and
(\ref{g0if}):  $\psi_1$, $\psi_2$, $X(R^2)$ and $Z_i$.  These
potentials are likely to be small, due to suppressing factors of
$c_2/\vert \mathbf{x}-\mathbf{x}^\prime \vert^2$.  The $\psi_1$ term
is somewhat promising as its suppression is polynomial, and not
exponential.  The $Z_i$ term is also interesting as it is has
polynomial amplification, as well as exponential suppression.  Such a
term may be potentially observable in experiments that measure vector
perturbations, such as gravity probe B \cite{gpb}.  Finally, we will
mention that the term proportional to $X(R^2)$ allows for the
possibility of constraining any $R^3$ term that may exist in the
generating Lagrangian.  Unlike the $R^2$ term, the $R^3$ term does not
first appear in the perturbative expansion as a Yukawa potential.  In
fact, in the limit that the $R^2$ term vanishes the potential $X(R^2)$
reduces to $\rho^2$, which could be observable in high
density environments.

\appendix

\section{Solving the $t-t$ equation to $O(4)$}
\label{appendixA}

In order to determine the $h_{00}^{(4)}$ term, it will first be
necessary to determine the Ricci scalar to $O(4)$.  This will be
achieved by solving the trace equation to the appropriate order.  In
the gauge defined by equations (\ref{gauge1}) and (\ref{gauge2}), the
trace of the field equations (\ref{field}) becomes
\begin{equation*}
\nabla^2 R^{(4)} - \frac{c_1}{6 c_2} R^{(4)} - R_{,00} + \frac{3
  c_3}{2 c_2} \nabla^2 R^2 + \frac{2}{c_1} U \nabla^2 R - 2
  \frac{c_2}{c_1} R \nabla^2 R - 24 \frac{c_2^2}{c_1^2} R_{,ij}
  R_{,ij} + 8 \frac{c_2}{c_1^2} U_{,ij} R_{,ij} = \frac{4 \pi}{c_2} p
  - \frac{4 \pi}{3 c_2} \rho \Pi
\end{equation*}
where here we have dropped the superscript on $R^{(2)}$, so that the $R$
above should be implicitly assumed to be of $O(2)$ (as can be
recognised from the required order of each term).  We now
introduce the new potentials
\begin{align}
\label{chi}
\hat{\chi} &\equiv \int \rho^\prime e^{-\sqrt{\frac{c_1}{6 c_2}} \vert
  \bf{x}-\bf{x}^\prime \vert} d^3x^\prime\\
\label{WX}
X(Q) &\equiv \int \frac{Q^\prime e^{-\sqrt{\frac{c_1}{6 c_2}} \vert
  \bf{x}-\bf{x}^\prime \vert}}{\vert \mathbf{x}-\mathbf{x}^\prime
  \vert} d^3x^\prime,
\end{align}
where a prime now denotes a quantity that is a function of $x^\prime$,
so that
\begin{align*}
\left( \nabla^2 - \frac{c_1}{6 c_2}\right)& \hat{\chi} = - \sqrt{6 c_1
c_2} R\\
\left( \nabla^2 - \frac{c_1}{6 c_2}\right)& X(Q) = - 4 \pi Q.
\end{align*}
Recognising the relations
\begin{align*}
U_{,ij} R_{,ij} &= \frac{1}{2} \left( \nabla^2 - \frac{c_1}{6
  c_2}\right) \nabla U \cdot \nabla R + \frac{2 \pi}{3 c_2} \nabla
  \rho \cdot \nabla U + 2 \pi \nabla \rho \cdot \nabla R\\
R_{,ij} R_{,ij} &= \left( \nabla^2 - \frac{c_1}{6 c_2}\right) \left(
  \frac{\vert \nabla R \vert^2}{2} - \frac{c_1 R^2}{24 c_2} \right)
  +\frac{4 \pi}{3 c_2} \nabla \rho \cdot \nabla R + \frac{c_1^2}{144
  c_2^2} R^2 - \frac{\pi c_1}{9 c_2^2} \rho R
\end{align*}
then allows the trace equation above to be integrated, to give
\begin{align}
\nonumber
R^{(4)} = &-\frac{\hat{\chi}_{,00}}{\sqrt{6 c_1 c_2}} - \left( \frac{3
c_3}{2c_2} + \frac{c_2}{c_1} \right)R^2 +12 \frac{c_2^2}{c_1^2} \vert
\nabla R \vert^2 - 4 \frac{c_2}{c_1^2} \nabla U \cdot \nabla R 
+ \frac{X(UR)}{12 \pi c_2}- \left(1
-\frac{c_1 c_3}{2 c_2^2} \right) \frac{X(R^2)}{8 \pi} \nonumber \\ &- \frac{2 X(\rho U)}{3 c_1
  c_2} + \frac{4 X( \rho R)}{3 c_1} + \frac{4 X(\nabla \rho \cdot \nabla U)}{3 c_1^2}
 \label{R4} - \frac{4 c_2 X(\nabla \rho
\cdot \nabla R)}{c_1^2} -\frac{X(p)}{c_2} + \frac{X(\rho \Pi)}{3 c_2}.
\end{align}
We are now sufficiently equipped to solve the $t-t$ field equation to
$O(4)$, in order to obtain $h_{00}^{(4)}$.  Equations (\ref{field})
give this as
\begin{multline*}
-\frac{c_1}{2} \nabla^2 h^{(4)}_{00} -\frac{1}{c_1} \nabla^2 U^2 +
 \left( \frac{3}{2} c_3- 2 \frac{c_2^2}{c_1} \right) \nabla^2 R^2 - 3
 \frac{c_2}{c_1} \nabla^2 U R -18 \frac{c_2^2}{c_1} \nabla^2 R_{,00} \\+
 6 \frac{c_2}{c_1} \nabla^2 U_{,00} +c_2 \nabla^2 R^{(4)} + 8 \frac{c_2}{c_1^2} U_{,ij}
 U_{,ij} -8 \frac{c_2^2}{c_1^2} U_{,ij} R_{,ij} - 48
 \frac{c_2^3}{c_1^2} R_{,ij} R_{,ij} \\+ \frac{5}{6} U R - \frac{c_2}{6}
 R^2 -  \frac{44\pi}{3 c_1} \rho U + \frac{52 \pi c_2}{3 c_1} \rho R =
 4 \pi \rho \Pi + 8 \pi \rho v^2 +12 \pi p
\end{multline*}
where the trace of (\ref{field}) has been used to eliminate the term
proportional to $R^{(4)}$.  On recognising
\begin{align*}
U_{,ij} U_{,ij} &= \frac{1}{2} \nabla^2 \vert \nabla U \vert^2 + 4\pi
\nabla \rho \cdot \nabla U\\
U_{,ij} R_{,ij} &= \frac{1}{2} \nabla^2 (\nabla U \cdot \nabla R) -
\frac{c_1}{24 c_2} \nabla^2 U R + \frac{2 \pi}{3 c_2} \nabla
  \rho \cdot \nabla U + 2 \pi \nabla \rho \cdot \nabla R -\frac{\pi
    c_1}{6 c_2} \rho R -\frac{\pi c_1}{18 c_2^2} \rho U
  +\frac{c_1^2}{144 c_2^2} U R \\
R_{,ij} R_{,ij} &= \frac{1}{2} \nabla^2 \vert \nabla R \vert^2 -
\frac{c_1}{12 c_2} \nabla^2 R^2 + \frac{4 \pi}{3 c_2} \nabla \rho
\cdot \nabla R +\frac{c1^2}{36 c_2^2} R^2 -\frac{2 \pi c_1}{9 c_2^2}
\rho R 
\end{align*}
this can be integrated to
\begin{align}
\nonumber
h^{(4)}_{00} = &-\frac{2}{c_1^2} U^2 + \left(3 \frac{c_3}{c_1}+4
\frac{c_2^2}{c_1^2} \right) R^2 - \frac{16 c_2}{3 c_1^2} U R - 36
\frac{c_2^2}{c_1^2} R_{,00} +12 \frac{c_2}{c_1^2} U_{,00}
+ 8 \frac{c_2}{c_1^3} \vert \nabla U \vert^2 - 48 \frac{c_2^3}{c_1^3}
\vert \nabla R \vert^2 \nonumber \\ &- 8 \frac{c_2^2}{c_1^3} \nabla U \cdot \nabla R
- \frac{7}{18 \pi c_1} \mathcal{V}(UR) + \frac{3 c_2}{4 \pi c_1} \mathcal{V}(R^2) +
\frac{64}{9 c_1^2} \mathcal{V}(\rho U) - \frac{44 c_2}{3 c_1^2} \mathcal{V}( \rho R)-
\frac{40 c_2}{3 c_1^3} \mathcal{V}(\nabla \rho \cdot \nabla U)
\nonumber \\ \label{h004a} &+ \frac{40
  c_2^2}{c_1^3} \mathcal{V}(\nabla \rho \cdot \nabla R) + \frac{2}{c_1} \mathcal{V}(\rho
\Pi) + \frac{4}{c_1} \mathcal{V}(\rho v^2) + \frac{6}{c_1} \mathcal{V}(p) + 2
\frac{c_2}{c_1} R^{(4)}
\end{align}
where
\begin{equation}
\label{VX}
\mathcal{V}(Q) \equiv \int \frac{Q^\prime}{\vert \mathbf{x}-\mathbf{x^\prime}
  \vert} dx^{\prime 3}.
\end{equation}
Equations (\ref{R4}) and (\ref{h004a}) now specify $h_{00}$ to $O(4)$.

\section{Re-expressing the potentials}
\label{appendixB}

The potentials $\mathcal{V}(\nabla \rho \cdot \nabla U)$,
$\mathcal{V}(\nabla \rho \cdot \nabla R)$, $X(\nabla \rho
\cdot \nabla U)$ and $X(\nabla \rho \cdot \nabla R)$ in equation (\ref{g00}) are not in usual
PPN form, as they are written as functions of gradients of $\rho$, $U$
and $R$.  Here we re-express these terms, with the gradients absent,
as
\begin{align*}
\mathcal{V}(\nabla \rho \cdot \nabla U) &= 4 \pi \mathcal{V}(\rho^2) +
\psi_1\\
\mathcal{V}(\nabla \rho \cdot \nabla R) &= \frac{4 \pi}{3 c_2}
\mathcal{V}(\rho^2) - \frac{c_1}{6 c_2} \mathcal{V}(\rho R) +
\sqrt{\frac{c_1}{6 c_2}} \frac{\psi_2}{3 c_2}+ \frac{\psi_3}{3 c_2} \\
X(\nabla \rho \cdot \nabla U) &= 4 \pi X(\rho^2) + \sqrt{\frac{c_1}{6
    c_2}} \psi_4 +\psi_5\\
X(\nabla \rho \cdot \nabla R) &= \frac{4 \pi}{3 c_2} X(\rho^2) -
\frac{c_1}{6 c_2} X(\rho R) +\frac{\psi_6}{3c_2} +\sqrt{\frac{c_1}{6c_2}}
\frac{(\psi_7+\psi_8)}{3c_2} + \frac{c_1 \psi_9}{18 c_2^2}.
\end{align*}
On substituting these expressions back into (\ref{g00}) the terms
proportional to $\mathcal{V}(\rho^2)$ and $X(\rho^2)$ cancel exactly.
The new potentials, $\psi_i$, are defined by
\allowdisplaybreaks
\begin{align*}
\psi_1 &\equiv \int \frac{\rho^\prime \rho^{\prime \prime} (\mathbf{x}-\mathbf{x}^\prime) \cdot
  (\mathbf{x}^\prime-\mathbf{x}^{\prime\prime})}{\vert
  \mathbf{x}-\mathbf{x}^\prime \vert^3 \vert
  \mathbf{x}^\prime-\mathbf{x}^{\prime\prime} \vert^3} d^3x^\prime
  d^3x^{\prime \prime}\\
\psi_2 &\equiv \int \frac{\rho^\prime \rho^{\prime \prime} (\mathbf{x}-\mathbf{x}^\prime) \cdot
  (\mathbf{x}^\prime-\mathbf{x}^{\prime\prime})}{\vert
  \mathbf{x}-\mathbf{x}^\prime \vert^3 \vert
  \mathbf{x}^\prime-\mathbf{x}^{\prime\prime} \vert^2} e^{-\sqrt{\frac{c_1}{6 c_2}} \vert
  \bf{x}^\prime-\bf{x}^{\prime\prime} \vert}
d^3x^\prime d^3x^{\prime \prime}\\
\psi_3 &\equiv \int \frac{\rho^\prime \rho^{\prime \prime} (\mathbf{x}-\mathbf{x}^\prime) \cdot
  (\mathbf{x}^\prime-\mathbf{x}^{\prime\prime})}{\vert
  \mathbf{x}-\mathbf{x}^\prime \vert^3 \vert
  \mathbf{x}^\prime-\mathbf{x}^{\prime\prime} \vert^3} e^{-\sqrt{\frac{c_1}{6 c_2}} \vert
  \bf{x}^\prime-\bf{x}^{\prime\prime} \vert}
d^3x^\prime d^3x^{\prime \prime}\\
\psi_4 &\equiv \int \frac{\rho^\prime \rho^{\prime \prime} (\mathbf{x}-\mathbf{x}^\prime) \cdot
  (\mathbf{x}^\prime-\mathbf{x}^{\prime\prime})}{\vert
  \mathbf{x}-\mathbf{x}^\prime \vert^2 \vert
  \mathbf{x}^\prime-\mathbf{x}^{\prime\prime} \vert^3} e^{-\sqrt{\frac{c_1}{6 c_2}} \vert
  \bf{x}-\bf{x}^{\prime} \vert}
d^3x^\prime d^3x^{\prime \prime}\\
\psi_5 &\equiv \int \frac{\rho^\prime \rho^{\prime \prime} (\mathbf{x}-\mathbf{x}^\prime) \cdot
  (\mathbf{x}^\prime-\mathbf{x}^{\prime\prime})}{\vert
  \mathbf{x}-\mathbf{x}^\prime \vert^3 \vert
  \mathbf{x}^\prime-\mathbf{x}^{\prime\prime} \vert^3} e^{-\sqrt{\frac{c_1}{6 c_2}} \vert
  \bf{x}-\bf{x}^{\prime} \vert}
d^3x^\prime d^3x^{\prime \prime}\\
\psi_6 &\equiv \int \frac{\rho^\prime \rho^{\prime \prime} (\mathbf{x}-\mathbf{x}^\prime) \cdot
  (\mathbf{x}^\prime-\mathbf{x}^{\prime\prime})}{\vert
  \mathbf{x}-\mathbf{x}^\prime \vert^3 \vert
  \mathbf{x}^\prime-\mathbf{x}^{\prime\prime} \vert^3} e^{-\sqrt{\frac{c_1}{6 c_2}} \vert
  \mathbf{x}-\mathbf{x}^{\prime} \vert -\sqrt{\frac{c_1}{6 c_2}} \vert
  \bf{x}^\prime-\bf{x}^{\prime\prime} \vert}
d^3x^\prime d^3x^{\prime \prime}\\
\psi_7 &\equiv \int \frac{\rho^\prime \rho^{\prime \prime} (\mathbf{x}-\mathbf{x}^\prime) \cdot
  (\mathbf{x}^\prime-\mathbf{x}^{\prime\prime})}{\vert
  \mathbf{x}-\mathbf{x}^\prime \vert^3 \vert
  \mathbf{x}^\prime-\mathbf{x}^{\prime\prime} \vert^2} e^{-\sqrt{\frac{c_1}{6 c_2}} \vert
  \mathbf{x}-\mathbf{x}^{\prime} \vert -\sqrt{\frac{c_1}{6 c_2}} \vert
  \bf{x}^\prime-\bf{x}^{\prime\prime} \vert}
d^3x^\prime d^3x^{\prime \prime}\\
\psi_8 &\equiv \int \frac{\rho^\prime \rho^{\prime \prime} (\mathbf{x}-\mathbf{x}^\prime) \cdot
  (\mathbf{x}^\prime-\mathbf{x}^{\prime\prime})}{\vert
  \mathbf{x}-\mathbf{x}^\prime \vert^2 \vert
  \mathbf{x}^\prime-\mathbf{x}^{\prime\prime} \vert^3} e^{-\sqrt{\frac{c_1}{6 c_2}} \vert
  \mathbf{x}-\mathbf{x}^{\prime} \vert -\sqrt{\frac{c_1}{6 c_2}} \vert
  \bf{x}^\prime-\bf{x}^{\prime\prime} \vert}
d^3x^\prime d^3x^{\prime \prime}\\
\psi_9 &\equiv \int \frac{\rho^\prime \rho^{\prime \prime} (\mathbf{x}-\mathbf{x}^\prime) \cdot
  (\mathbf{x}^\prime-\mathbf{x}^{\prime\prime})}{\vert
  \mathbf{x}-\mathbf{x}^\prime \vert^2 \vert
  \mathbf{x}^\prime-\mathbf{x}^{\prime\prime} \vert^2} e^{-\sqrt{\frac{c_1}{6 c_2}} \vert
  \mathbf{x}-\mathbf{x}^{\prime} \vert -\sqrt{\frac{c_1}{6 c_2}} \vert
  \bf{x}^\prime-\bf{x}^{\prime\prime} \vert}
d^3x^\prime d^3x^{\prime \prime}.
\end{align*}

\end{document}